# Para-B&B: Load-Balanced Deterministic Parallelization of Solving MIP

**Jinyu Zhang, Di Huang\*, Yue Liu, Shuo Wang, Zhenyu Pu, Zhiyuan Liu**

School of Transportation, Southeast University, Nanjing, China
jinyuzhang@seu.edu.cn, dihuang@seu.edu.cn, yueliu80806@163.com, wangshuo@seu.edu.cn,
220245416@seu.edu.cn, zhiyuanl@seu.edu.cn

**Abstract**

Mixed-integer programming (MIP) extends linear programming by incorporating both continuous and integer decision variables, making it widely used in production planning, logistics scheduling, and resource allocation. However, MIP remains NP-hard and cannot generally be solved to optimality in polynomial time. Branch-and-bound, a fundamental exact method, faces significant parallelization challenges due to computational heterogeneity and strict determinism requirements in commercial applications. This paper presents the first fully open-source implementation of deterministic parallel branch-and-bound for HiGHS, a high-performance MIP solver. Our approach introduces a novel data-parallel architecture ensuring strict determinism by replicating complete solver state across worker threads and eliminating non-deterministic synchronization primitives. A key innovation is our AI-driven load balancing mechanism employing multi-stage workload prediction models that estimate node computational complexity based on structural characteristics and historical performance data, coupled with dynamic parameter adjustment strategies. The framework executes orchestrated parallel phases including concurrent dive operations, systematic data consolidation, and intelligent node selection. Comprehensive experimental evaluation on 80 MIPLIB 2017 benchmark instances demonstrates effectiveness, achieving a geometric mean speedup of 2.17 using eight threads while maintaining complete deterministic guarantees. Performance gains become increasingly pronounced for higher node counts, with speedup factors reaching 5.12 for computationally intensive instances and thread idle rates averaging 34.7%.

## 1 Introduction

Mixed-integer programming (MIP) is a cornerstone of operations research, discrete optimization, and artificial intelligence, with growing complexity driven by applications in logistics, finance, energy systems, and machine learning (Bertsimas and Tsitsiklis, 1997). Despite major advances in MIP solvers, effectively parallelizing branch-and-bound (B&B) remains a central challenge that limits scalability on multicore architectures.

Parallelizing B&B presents unique difficulties because branching decisions, domain propagations, and solution quality are tightly coupled, making naive approaches inadequate. Commercial applications often require strict determinism—identical inputs must yield identical outputs regardless of parallel execution parameters. This requirement excludes techniques that rely on race conditions or timing-dependent behavior, creating fundamental tension between speedup and reproducibility.

The current MIP solver landscape is sharply divided between commercial and open-source offerings. Commercial solvers, including Gurobi, CPLEX, and Xpress, typically outperform open-source counterparts by one to two orders of magnitude. Among open-source solvers, SCIP represents the most mature platform with established parallel extensions (Berthold et al., 2009). ParaSCIP provides distributed memory parallelization for massively parallel environments (Shinano et al., 2011), while FiberSCIP offers shared memory parallelization for multi-threaded computation (Shinano et al., 2018). However, both extensions depend on the proprietary UG (Ubiquity Generator) framework, a generalized abstraction layer designed to parallelize any state-of-the-art MIP solver across different computational environments, whose source code is not publicly available. This proprietary nature significantly limits reproducibility, research adaptation, and community development of parallel B&B algorithms.

HiGHS emerges as an attractive platform for parallel development due to several distinctive characteristics. As a high-performance open-source solver, HiGHS demonstrates exceptional serial performance that even surpasses 8-core parallel SCIP on many instances. Its clean, thread-safe architecture facilitates development and modification, while its efficient handling of large-scale sparse problems and strong academic and industrial adoption (Huangfu and Hall, 2018) establish it as a solid foundation for research innovation. However, HiGHS currently lacks comprehensive parallel branch-and-bound implementation, presenting a critical gap in the open-source ecosystem where no fully accessible deterministic parallel solver exists.

This work addresses this fundamental gap by implementing the first publicly available deterministic parallel branch-and-bound algorithm for HiGHS. Our framework ensures strict deterministic behavior through innovative data paral-lelization strategies that eliminate race conditions while incorporating intelligent load balancing mechanisms. We introduce a novel data-parallel architecture that replicates global solver state across worker threads, enabling independent computation phases with deterministic synchronization.

---

\*Corresponding author

The contributions are threefold. First, we develop a comprehensive data parallelization strategy that addresses race conditions inherent in parallel branch-and-bound by replicating contested data structures across workers and employing deterministic merging and broadcasting protocols, effectively eliminating non-deterministic synchronization primitives. Second, we introduce a machine-learning-driven load balancing framework that predicts node computational complexity and dynamically adjusts worker parameters, combining multi-stage workload prediction with adaptive parameter tuning to address computational heterogeneity in tree exploration. Third, we provide the first complete open-source implementation of deterministic parallel branch-and-bound for MIP, with full source code availability enabling reproducible research and community adoption.

The remainder of this study is organized as follows. Section 2 reviews related work in parallel B&B algorithms and load balancing strategies. Section 3 establishes mathematical and algorithmic foundations. Section 4 presents the sequential HiGHS algorithm foundation. Section 5 and Section 6 describe our deterministic parallel framework. Section 7 introduces our machine-learning-driven load balancing approach. Section 8 presents comprehensive experimental results demonstrating effectiveness across diverse problem instances.

## 2 Related Works

### 2.1 Parallel branch-and-bound algorithm

Parallelizing B&B is inherently challenging because tree search is fundamentally sequential, and branching decisions are tightly coupled with domain propagation (Schryen, 2020). Foundational work by Gendron and Crainic (1994) establishes a systematic taxonomy of parallel B&B schemes with respect to coordination mechanisms and degree of parallelism.

Building on this foundation, Ralphs et al. (2018) formalize four canonical levels of parallelism: tree, subtree, node, and subnode. Tree parallelism launches multiple, differently guided searches in parallel, with beneficial information (e.g., solutions and bounds) shared across searches. Subtree parallelism distributes independent subproblems to serial solvers for time-limited exploration, avoiding finer-grained synchronization overhead. Node parallelism, the most widely adopted approach, processes multiple nodes concurrently under centralized control, typically through master-worker architectures where masters coordinate task distribution and workers execute node computations (Lei et al., 2024). Subnode parallelism parallelizes within-node components (e.g., heuristics and domain propagation), though strong data dependencies often limit its effectiveness.

### 2.2 Load balancing strategies

Load balancing in parallel B&B presents significant challenges due to irregular, path-dependent search trees that generate unpredictable workload variations across threads. Node processing costs vary dramatically with problem structure and evolve dynamically (Schryen, 2020).

Static approaches distribute initial work to minimize processor idle time through four schemes: root initialization (distributing root children iteratively), enumerative initialization (replicating root problems), selective initialization (generating single paths), and direct initialization (creating nodes at specific depths) (Ralphs et al., 2018).

Dynamic approaches become essential when static seeding fails due to uneven utilization and evolving tree structures, employing centralized work sharing (manager-coordinated transfers) or decentralized work stealing (peer-to-peer task requests).

Effective load balancing must balance work quality (node priority based on bounds) against work quantity (available task volume) while maintaining high processor utilization. Contemporary AI-driven frameworks extract features including branch depth, LP gaps, and iteration counts to predict workloads and detect anomalies, integrating global queue management with worker-specific selection rules.

However, dynamic load balancing strategies, while computationally more efficient, fundamentally compromise deterministic behavior. The timing-dependent nature of work stealing and adaptive task distribution creates non-reproducible execution paths, where identical problem instances may explore different search trajectories across multiple runs, violating the determinism requirements essential for commercial applications.

### 2.3 Parallel solvers

The parallel MIP solver landscape is sharply divided between commercial and open-source offerings. Commercial solvers achieve substantial performance gains through sophisticated parallel implementations, whereas open-source solvers have historically been largely limited to sequential execution (Ralphs et al., 2018). Table 1 summarizes the current landscape of parallel solvers and their key characteristics.

Commercial solvers emphasize shared-memory parallelism optimized for multicore hardware. CPLEX provides comprehensive parallel functionality across LP optimization, B&B tree search, and heuristics, with Fischetti et al. (2016)

demonstrating consistent speedups through randomized parallel sampling on non-trivial instances. Gurobi employs multi-threaded architectures with intelligent load balancing and work distribution while offering deterministic modes that guarantee reproducible results. Xpress parallelizes multiple algorithmic components—tree search, LP solving, heuristics, and cut generation—achieving determinism through synchronized thread execution (Berthold et al., 2018).

Open-source development has followed diverse trajectories, with SCIP representing the most mature platform. Two notable SCIP extensions exemplify complementary strategies: ParaSCIP enables distributed-memory parallelism via the UG framework, scaling to 80,000 cores on supercomputers for challenging MIPLIB instances (Shinano et al., 2016), while FiberSCIP provides shared-memory parallelism through dynamic search tree partitioning across SCIP instances (Shinano et al., 2018). Specialized solvers like PIPS-SBB target dual-block-angular structured problems with multilevel parallelism for stochastic MIPs (Munguía et al., 2019), while emerging GPU-based approaches such as FastDOG demonstrate massively parallel implementations for specific algorithmic components (Abbas & Swoboda, 2022). HiGHS represents a compelling opportunity for advancing open-source parallel capabilities. Despite its exceptional performance, broad adoption, and proven reliability, no comprehensive parallel B&B implementation exists for HiGHS.

## 3 Preliminaries

This section sets out the mathematical foundations and algorithmic components underlying our parallel B&B framework. We first formalize the structure of the MIP, then provide a comprehensive overview of the B&B algorithm and its critical computational modules.

**Mixed-integer programming.** A mixed integer programming problem seeks to optimize a linear objective function subject to linear constraints where some variables are restricted to integer values. The standard formulation can be expressed as:

$$\min_x c^T x \quad (1)$$

s.t.
$$Ax \leq b, \quad (2)$$
$$l \leq x \leq u, \quad (3)$$
$$x_j \in \mathbb{Z}, \forall j \in I, \quad (4)$$

where $c \in \mathbb{R}^n$ represents the objective coefficients, $A \in \mathbb{R}^{m \times n}$ is the constraints matrix, $b \in \mathbb{R}^m$ contains the right-hand side values, and $l$, $u$ define variable bounds. The index set $I \subseteq \{1, 2, \ldots, n\}$ identifies integer-constrained variables, while the remaining variables are continuous.

**Branch-and-bound (B&B) algorithm.** The B&B algorithm systematically explores the solution space by recursively partitioning the feasible region and computing bounds on the optimal objective value. It maintains a search tree in which each node defines a subproblem obtained from the original MIP by fixing or restricting a subset of variables. The root node corresponds to the linear programming (LP) relaxation formed by dropping integrality constraints. At each node, the LP relaxation is solved to obtain a bound on the best attainable objective in that subregion (a lower bound for minimization). B&B advances via three core operations: **branching**, which creates child subproblems typically by imposing complementary bounds on a fractional integer variable; **bounding**, which computes tight bounds for each subproblem via the LP relaxation; and **pruning** (fathoming), which discards subproblems that cannot yield an improving integer solution because they are infeasible, their bound is no better than the incumbent upper bound, or they produce an integer solution that does not improve the incumbent.

Table 1: Overview of Parallel Solver Implementations.

| Solver | Type | Parallel Method | Memory Model | Deterministic | Open Source | Key Features |
|---|---|---|---|---|---|---|
| CPLEX | Commercial | Multi-threaded | Shared | Yes | No | Randomized sampling, concurrent optimization |
| Gurobi | Commercial | Multi-threaded | Shared | Yes | No | Advanced load balancing, intelligent distribution |
| Xpress | Commercial | Multi-threaded | Shared | Yes | No | Multi-component parallelization, synchronized execution |
| ParaSCIP | Academic | MPI | Distributed | Yes | Framework only | Massive scalability (80K cores), UG framework |
| FiberSCIP | Academic | Pthreads | Shared | Yes | Framework only | External tree splitting, multiple SCIP instances |
| PIPS-SBB | Academic | MPI | Distributed | Yes | Limited | Structured problems, multi-level parallelism |
| Para-B&B (this study) | Open-source | Multi-threaded | Shared | Yes | Yes | Data parallelization, ML-driven load balancing |

**Domain propagation**. Domain propagation tightens variable bounds by exploiting constraint structure and logical implications. The propagation engine iteratively analyzes constraints to derive bound improvements, revisiting affected constraints until a fixed point ensures global consistency. Specialized routines handle bound tightening for linear constraints and implication-graph analysis. Effective propagation reduces subproblem complexity, strengthens relaxations, and enables early infeasibility detection.

**Separation**. Cutting planes are linear inequalities valid for integer-feasible solutions but violated by fractional LP solutions. Modern MIP solvers systematically identify violated valid inequalities and add them to strengthen relaxations and reduce integrality gaps. Multiple cut families include Gomory cuts, cover inequalities, and problem-specific cuts. Separation routines analyze fractional solutions while cut-management policies balance relaxation strength against computational cost using cut pools and aging rules.

**Conflict analysis**. Conflict analysis learns from infeasible subproblems by deriving constraints that preclude re-exploration of analogous regions. When propagation detects infeasibility, the module backtracks through the implication graph to identify minimal assignments causing contradiction, yielding learned conflict constraints. A global conflict pool manages these constraints with aging policies and applies them across nodes to strengthen pruning throughout the search.

**Primal heuristics**. Primal heuristics complement systematic B&B by rapidly producing high-quality feasible incumbents from partial assignments, fractional solutions, or related instances. Key components include random rounding, RENS, and RINS. Effective scheduling triggers heuristics based on search state and resource availability, with strong incumbents enabling aggressive pruning and faster convergence.

## 4 HiGHS Sequential Algorithm

This section describes the sequential MIP algorithm implemented in HiGHS, which underpins our parallel framework. The HiGHS sequential MIP solver follows a classical B&B framework augmented with modern algorithmic components. It proceeds through a staged workflow that reduces model complexity then explores the search space. The process begins with initialization and intensive presolve, applying constraint elimination, variable-bound tightening, and structural reformulations to substantially shrink the model.

After presolve, the algorithm performs root-node processing, solving the LP relaxation to establish initial bounds, performing cut separation, and invoking primal heuristics to obtain incumbents. It manages the cut pool, aging inequalities that haven't contributed to bound improvement.

The core B&B tree search explores the solution space by recursively partitioning and computing bounds. The solver maintains a priority queue of active nodes with selection policies blending best-bound guidance and numerical stability. Search alternates between intensive local exploration via diving and global coordination through bound updates and queue management.

At each iteration, the algorithm selects a promising node and initiates a dive for local exploration. The dive proceeds through node evaluation, branching decisions, and bound strengthening until stopping criteria are met. Between dives, global coordination propagates bound improvements, prunes infeasible regions, and updates search structures.

The algorithm employs adaptive mechanisms responding to instance characteristics, including dynamic node-selection policies, adaptive iteration limits, and restart strategies. Upon termination, the solver performs cleanup and produces a solution report.

This sequential structure underpins our parallel design. The central insight is that dive operations and node-selection can be executed concurrently on disjoint tree nodes, whereas global coordination phases require careful synchronization to preserve correctness and determinism.

## 5 Parallel solving root node

Root-node parallelization in MIP solvers has received comparatively little attention, both in practice and in the literature. CPLEX offers deterministic and opportunistic parallel modes for root-node processing, yet there is little published work on optimizing these modes specifically for the root phase. Gurobi's Concurrent Optimizer enables parallel exploration but has not been extensively studied in the context of root-node parallelization. SCIP implements some parallel features at the root, but development has largely focused on branch-and-bound tree search rather than the root phase itself. Overall, academic work on root-node parallelization remains sparse, with most efforts concentrated on the subsequent tree-search stage.

As discussed in Section 4, root-node processing is pivotal in MIP solving. Parallelizing selected root-node workflows can materially improve overall efficiency, particularly for instances resolved entirely at the root. Our analysis of HiGHS's root-node pipeline identified several components amenable to parallel execution. We therefore implemented targeted parallel strategies in these critical stages to enhance computational efficiency and better utilize CPU threads.

Root-node processing in MIP comprises two primary components: (i) cut generation (separation), which tightens the LP relaxation by adding valid inequalities to improve the lower bound and shrink the feasible region; and (ii) primal heuristics, which seek feasible solutions to establish upper bounds and guide optimality assessment.

Additionally, root-node evaluation includes symmetry detection to identify structural regularities and analytic-center computation to provide geometric information that guides heuristics. HiGHS parallelizes both components: symmetry detection runs asynchronously on a background thread, and analytic-center computation executes in a separate thread via an independent IPM solver instance; both operations are synchronized during the separation phase.

Building on the cut-generation and heuristic components described above, we identify several additional parallelizable submodules and design tailored schemes for each. In the separation phase, multiple separators are invoked concurrently to accelerate the discovery of effective cutting planes and tighten the relaxation more rapidly. In the heuristic phase, we parallelize selected steps of Randomized Rounding and RENS (Relaxation Enforced Neighborhood Search), enabling the simultaneous generation of multiple candidate incumbents and thereby improving upper-bound quality.

# 6 Parallel branch-and-bound algorithm

Determinism is crucial in parallel branch-and-bound: it ensures reproducibility and enables consistent evaluation of algorithmic performance. Advanced commercial solvers, e.g., Gurobi, COPT, and Optverse, provide deterministic parallel modes. Broadly, parallelization strategies fall into two classes: lock-based synchronization and data parallelism. Lock-based approaches often introduce timing-dependent behavior due to thread scheduling, undermining determinism. To avoid this, our framework adopts a data-parallel design: we replicate potentially contended state into identical worker-local copies, update them in parallel, and then reconcile changes via a barrier-synchronized, deterministic merge-and-broadcast step. This eliminates race conditions and guarantees run-to-run deterministic results.

**Framework**. The proposed parallel B&B algorithm adopts a master-worker architecture (see Figure 2). The master thread maintains the global state, managing the node queue, updating shared information, and coordinating synchronization, while worker threads independently execute compute-intensive tasks, notably parallel node selection and diving, using the state provided by the master.

**Data replication**. Each worker thread maintains a full replica of the global solver state, including variable domains, cut and conflict pools, the LP, pseudocosts, and related data structures. This replication obviates complex shared-memory synchronization that could introduce nondeterminism. With local copies, workers execute compute-intensive tasks independently, free from interference by concurrent threads. The design trades additional memory for computational independence and determinism, that is, a favorable exchange given the typically modest size of the solver state relative to the substantial performance gains from parallelism.

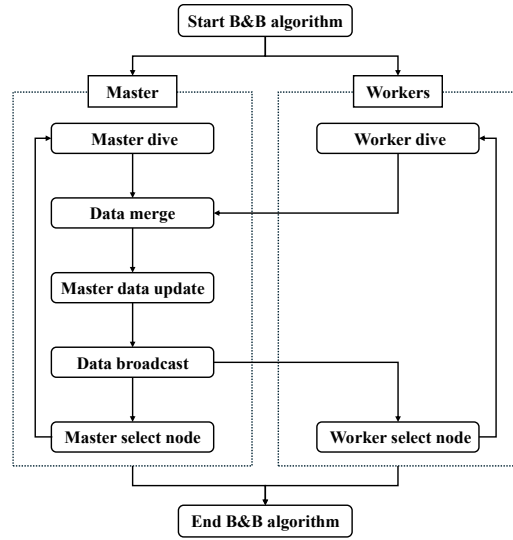

Figure 2: The framework of parallel B&B algorithm.

**Parallel dive phase**. During the dive phase, each worker conducts an independent depth-first search from a distinct node of the B&B tree, using its private replica of the solver state to perform node evaluation, constraint propagation, primal heuristics, and local tree expansion. The parallel diving scheme employs adaptive load balancing that dynamically tunes per-worker iteration budgets based on historical throughput and current computational demand. A worker continues its dive until predefined stopping criteria are met, for example, exhaustion of its iteration budget, a time limit, or completion of the assigned subtree.

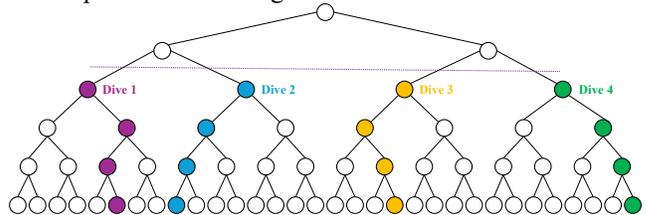

Figure 3: The illustration of parallel dive search.

**Data Consolidation**. After the parallel dive phase, per-worker findings and results are systematically consolidated into the global state. The consolidation uses specialized merge routines tailored to each data type, optimal solution transfer, domain-bound propagation, cut-pool integration, and statistics aggregation. To ensure determinism, worker contributions are processed in a fixed, predefined order, eliminating race conditions and yielding reproducible outcomes across runs.

**Global state synchronization**. The master thread updates the global solver state with the merged contributions from all workers. During this synchronization step, it performs global domain propagation, prunes infeasible nodes, tightens the global lower bound, and computes the optimality gap.

This stage serves as the central coordination point, unifying distributed computations into a coherent view of search progress. It also includes restart-detection logic that monitors efficiency and triggers strategic restarts when advantageous.

**Information broadcasting**. Updated global information is disseminated to all workers via a deterministic broadcast. This ensures that each worker begins the next iteration with an identical global state, including updated bounds, variable-domain restrictions, incumbent solutions, and search statistics. To limit communication overhead, the broadcast uses selective dissemination, prioritizing data that directly affects search efficiency; relevance filters transmit only the most useful cuts, conflicts, and heuristic artifacts.

**Parallel node selection**. Workers coordinate via a distributed node-selection stage to identify the most promising nodes for the next iteration. This phase couples global queue management with worker-local selection policies, enabling parallel priority evaluation while preserving overall search coherence. The implementation (see Figure 4) employs sophisticated load-distribution algorithms that jointly consider node promise and cross-worker balance. Robust fallback policies ensure reliable assignment under edge cases and transient imbalances.

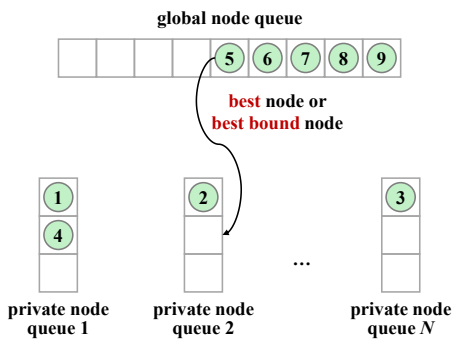

Figure 4: The illustration of node predistribution.

## 7 Machine-learning-driven load balance

Effective load balancing is central to parallel B&B algorithms due to irregular, path-dependent workload variations across threads. Static allocations are inadequate because node difficulty varies with problem structure and solver state. We address this with an AI-driven framework coupling intelligent dive cycle prediction with dynamic parameter tuning.

**Predictive dive cycle count estimation.** Our approach operates at individual dive granularity, representing each dive through features combining local signals (branching state, LP metrics) with global context (bounds, cuts, solver history). Key features include branch depth, search index, LP objective gap, iteration count, fixed/fractional variables, and parent-node statistics. These capture both structural characteristics and runtime behavior for effective load balancing.

Figure 4 reports absolute correlation magnitudes for the key features, underscoring the dominant role of LP metrics and bound information.

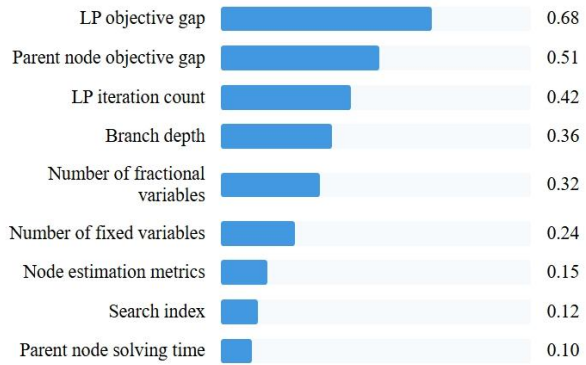

Figure 5: Feature importance.

**Multi-stage machine learning approach.** Our strategy employs a four-stage workflow:

*Stage 1:* Linear regression predicts computational intensity from structural features, establishing baseline workload distribution across workers.

*Stage 2*: Critical-node identification applies dual criteria - robust deviation tests flag nodes exceeding median intensity by three median absolute deviations, while cross-worker comparisons trigger rebalancing when aggregate burdens deviate from global means.

*Stage 3*: Enhanced prediction applies ridge regression and gradient-boosted decision trees for flagged critical nodes, capturing higher-order feature interactions and improving accuracy.

*Stage 4*: Model-quality monitoring compares predicted vs. actual execution times via Absolute Percentage Error. Systems with five consecutive predictions exceeding 50% APE trigger automatic retraining.

**Dynamic parameter-based load control**. Beyond task distribution, we dynamically tune dive parameters using execution-profile analysis. When imbalances are detected, overloaded threads receive less aggressive settings (higher thresholds, lower limits, looser tolerances) while underutilized threads get intensive configurations (tighter tolerances, deeper budgets, stricter criteria).

**Model selection**. Empirical evaluations show linear regression, exponential regression, and gradient boosting deliver optimal accuracy-inference trade-offs. The integrated approach yields self-adaptive load balancing that responds dynamically to evolving computational patterns, achieving substantial speedups over static allocation, especially on instances with high computational heterogeneity.

## 8 Numerical experiments

In this section, we present comprehensive experimental results demonstrating the effectiveness of our Para-B&B implementation. Our evaluation focuses on parallel scalability, load balancing efficiency, and deterministic behavior preservation across diverse problem instances.

All experiments are conducted on an Intel Core i5-12400H CPU at 2.5 GHz with 16 GB RAM. We select 80 instances from the MIPLIB 2017 benchmark suite following established criteria for evaluating parallel branch-and-bound performance. Instances are filtered to include only those solvable within a 7,200-second time limit using sequential HiGHS. This selection strategy ensures sufficient computational complexity to demonstrate parallel algorithm benefits while maintaining reasonable experimental execution times.

We employ multiple complementary metrics to provide comprehensive performance evaluation. **Speedup** is calculated as the ratio of sequential execution time to parallel execution time for identical problem instances:

$$\text{Speedup} = \frac{T_{serial}}{T_{parallel}}, \quad (5)$$

where $T_{serial}$ and $T_{parallel}$ represent the solving time of serial and parallel algorithms, respectively.

**Thread idle rate** measures load balancing effectiveness by quantifying the percentage of time worker threads spend waiting for new work assignments:

$$\text{Thread Idle Rate} = \frac{\sum_i W_i}{\sum_i E_i} \times 100\%, \quad (6)$$

where $W_i$ and $E_i$ represent the cumulative wait and execution time for thread $i$, respectively.

Tables 2 and 3 present detailed experimental results for 10 representative instances, including explored nodes, LP iterations, objective values, and timing information. These instances span different problem characteristics and computational complexities to provide comprehensive performance insights.

Table 2: Sequential Execution Results on Selected MIPLIB 2017 Instances.

| Case | objective  | timing  | nodes   | iterations |
|------|------------|---------|---------|------------|
| A    | -52301     | 40.83   | 13775   | 457235     |
| B    | 1          | 90.70   | 1002270 | 2290631    |
| C    | 40005.05414| 159.50  | 374840  | 2426251    |
| D    | 0          | 342.45  | 10736   | 722781     |
| E    | 11         | 352.88  | 380874  | 4013882    |
| F    | 3712       | 510.14  | 51930   | 5619804    |
| G    | 211913     | 634.85  | 24568   | 2790239    |
| H    | 1200012600 | 718.80  | 505194  | 11442089   |
| I    | 11801.18573| 1788.48 | 4396296 | 29743291   |
| J    | -52301     | 40.83   | 13775   | 457235     |

Table 3: Parallel Execution Results with Eight Threads.

| Case | timing | nodes  | iterations | Speedup |
|------|--------|--------|------------|---------|
| A    | 15.94  | 10660  | 388591     | 2.56    |
| B    | 32.36  | 756705 | 1698412    | 2.80    |
| C    | 39.54  | 353100 | 2000804    | 4.03    |
| D    | 144.29 | 7237   | 649419     | 2.37    |
| E    | 171.76 | 555576 | 4944462    | 2.05    |
| F    | 289.35 | 72740  | 9427207    | 1.76    |
| G    | 344.84 | 43785  | 4639923    | 1.84    |
| H    | 351.41 | 979714 | 8515188    | 2.05    |
| I    | 349.1  | 3182390| 20133421   | 5.12    |
| J    | 15.94  | 10660  | 388591     | 2.56    |

Across our test set of 80 instances, the eight-thread speedup distribution ranges from 1.00 to 5.12, with all instances achieving correct optimal solutions. The geometric mean of speedup is 2.17. The thread idle rate distribution spans from 18.5% to 56.6%, with an average of 34.7%. This indicates that approximately 30% of computational resources are underutilized due to load imbalance in our parallel algorithm, highlighting both the effectiveness of our approach and opportunities for further optimization.

To provide deeper insights into load balancing behavior, Tables 4 and 5 present detailed thread utilization statistics for two representative instances with contrasting computational characteristics. The analysis includes work time, wait time, dive operations count, and idle rate for each thread. The two instances exhibit markedly different computational characteristics that directly impact load balancing effectiveness. The lower absolute idle rates in brazil3 demonstrate that our load balancing mechanism performs more effectively on instances with sustained computational workloads, where the overhead of work redistribution is better amortized across longer processing times.

Table 4: Thread Utilization Analysis for pk1 Instance

| Thread   | Work/Wait Time (s) | Dives | Idle Rate (%) |
|----------|--------------------|-------|---------------|
| master   | 154.81/58.76       | 1742  | 27.51         |
| worker 1 | 165.12/48.45       | 1741  | 22.69         |
| worker 2 | 154.84/58.73       | 1741  | 27.5          |
| worker 3 | 148.44/65.13       | 1741  | 30.5          |
| worker 4 | 156.40/57.17       | 1741  | 26.77         |
| worker 5 | 154.56/59.01       | 1740  | 27.63         |
| worker 6 | 151.73/61.84       | 1740  | 28.96         |
| worker 7 | 161.28/52.29       | 1741  | 24.48         |

Table 5: Thread Utilization Analysis for brazil3 Instance.

| Thread   | Work/Wait Time (s) | Dives | Idle Rate (%) |
|----------|--------------------|-------|---------------|
| master   | 2411.40/601.31     | 21    | 19.96         |
| worker 1 | 2359.65/653.07     | 20    | 21.68         |
| worker 2 | 2395.56/617.15     | 20    | 20.48         |
| worker 3 | 2491.68/521.03     | 20    | 17.29         |
| worker 4 | 2541.59/471.13     | 20    | 15.64         |
| worker 5 | 2545.29/467.43     | 20    | 15.52         |
| worker 6 | 2417.46/595.26     | 20    | 19.76         |
| worker 7 | 2483.51/529.21     | 20    | 17.57         |

## 9 Conclusion

This work presents the first fully open-source implementation of deterministic parallel branch-and-bound for the HiGHS MIP solver, addressing a critical gap in the open-source optimization ecosystem. Our Para-B&B framework introduces innovative data parallelization strategies and AI-driven load balancing mechanisms that achieve substantial performance improvements while maintaining strict deterministic guarantees.

The data-parallel architecture developed in this study ensures strict determinism by replicating global solver state across worker threads and eliminating non-deterministic synchronization primitives. This approach fundamentally addresses race conditions inherent in parallel branch-and-bound algorithms through deterministic merging and broadcasting protocols. The integration of machine learning-driven load balancing represents a significant advancement in addressing computational heterogeneity, employing multi-stage workload prediction models that estimate node computational complexity based on structural characteristics and historical performance data.

Comprehensive experimental evaluation on 80 instances from the MIPLIB 2017 benchmark suite demonstrates the effectiveness of our approach, achieving a geometric mean speedup of 2.17 using eight threads while maintaining complete deterministic guarantees. The eight-thread speedup distribution ranges from 1.00 to 5.12, with all instances achieving correct optimal solutions. Performance gains become increasingly pronounced for instances with higher node counts, with speedup factors reaching 2.53 for instances exceeding 60,000 nodes. The framework demonstrates effective load balancing with thread idle rates averaging 34.7%, indicating efficient resource utilization across worker threads. The framework's scalability follows approximately linear growth, indicating strong potential for larger-scale parallel deployments.

Para-B&B fills a critical void in the open-source MIP solver landscape by providing the first complete, publicly available deterministic parallel implementation. Unlike existing parallel extensions that depend on proprietary frameworks, our solution offers full source code availability, enabling reproducible research and community-driven development. This accessibility democratizes access to high-performance parallel optimization capabilities and provides a solid foundation for future innovations in parallel mixed-integer programming, with promising implications for both academic research and industrial applications requiring strict determinism.

### A. HiGHS Serial B&B

**Algorithm 1: HiGHS Sequential MIP Solver**

1. Initialize MIP instance, set global lower bound to -∞, global upper bound to +∞, create priority queue and search stack
2. Presolve to reduce problem complexity
3. **If** problem solved during presolve then return solution
4. Apply trivial heuristics to find initial feasible solution
5. Evaluate root node: solve LP relaxation and perform cut generation
6. Perform aging to shrink cut pool after root separation
7. Update lower bound and upper bound from root node results
8. Add root node to priority queue
9. **While** priority queue is not empty **do**
10.     Perform aging on conflict pool
11.     Set iteration limit for LP solves during dive
12.     Choose best node from queue based on lower bound or estimation and install to search
13.     **While** search has active node **do**
14.         Evaluate node: solve LP relaxation
15.         **If** suboptimal (iteration limit reached) then add current node back to queue and **break**
16.         **If** node is pruned then increment leaf count and **break**
17.         **If** node not pruned and heuristics allowed then
18.             Apply primal heuristics (randomized rounding, RENS, RINS)
19.         **If** domain becomes infeasible then **break**
20.         **If** node not pruned then
21.             Dive: perform branching and select child node
22.             **If** suboptimal then **break**
23.             Increment leaf count
24.         **If** limits reached then set limit_reached and **break**
25.         **If** dive depth >= 100 nodes then **break**
26.         **If** cannot backtrack in current plunge then **break**
27.         Perform conflict pool aging **if** needed
28.     Add open nodes from search to priority queue
29.     **If** limit reached then update bounds and **break**
30.     Propagate global domain
31.     Prune infeasible nodes in queue based on global domain
32.     **If** global domain infeasible then clear queue and **break**
33.     Update global lower bound from queue
34.     **If** domain changes found then update local domain and cleanup fixed variables
35.     **If** restart conditions met then perform restart and go to step 4
36.     **While** queue not empty **do**
37.         Select next node from queue (best bound or best estimate)
38.         Install node to search
39.         Evaluate node: solve LP relaxation
40.         **If** suboptimal then add node back to queue
41.         **If** node pruned then
42.             Backtrack, increment counters, propagate domain, prune infeasible nodes
43.             Update bounds and **continue** to next node
44.         **If** node not pruned then
45.             Separate this node: generate cutting plane
46.             Store basis and **break** to perform dive
47. Clean up and return solution

**B. Para-B&B Deterministic Parallel B&B**

**Algorithm 2: HiGHS Parallel Solver with Load Balancing**

1. Initialize MIP instance, set global bounds, create worker threads
2. Presolve to reduce problem complexity
3. **If** problem solved during presolve then terminate workers and return solution
4. Apply trivial heuristics to find initial feasible solution
5. Evaluate root node: solve LP relaxation and perform cut generation
6. Perform aging to shrink cut pool after root separation
7. Create worker data replicas and broadcast global information to all workers
8. Initialize master search and install root node
9. **While** active workers exist do
10.     Perform aging on conflict pools for master and active workers
11.     Set adaptive iteration limits for master and workers based on historical performance
12.     Start parallel dive phase with synchronized timing
13.     Master and active workers perform concurrent dive operations:
14.         Evaluate current node and apply primal heuristics **if** allowed
15.         **If** node pruned then increment leaf count and **continue**
16.         **If** node not pruned then dive by branching and selecting child nodes
17.         **Continue** diving until depth limits, iteration limits, or backtrack conditions met
18.     Wait for all workers to complete dive phase using barrier synchronization
19.     Merge worker data to global state:
20.         Transfer incumbent solutions and update bounds
21.         Merge domain information and node queues
22.         Consolidate cut pools, conflict pools, and heuristic statistics
23.         Update global counters and performance metrics
24.     Update global information:
25.         Add open nodes from all searches to global queue
26.         Propagate global domain and prune infeasible nodes
27.         Update global bounds and check termination conditions
28.     **If** restart conditions met then perform restart and go to step 4
29.     Broadcast updated global information to all workers:
30.         Synchronize domain restrictions and bounds
31.         Transfer incumbent solutions and optimization limits
32.         Update worker-specific parameters and statistics
33.     Start parallel node selection phase:
34.         Distribute available nodes to private queues using round-robin allocation
35.         For each worker requiring a node do
36.             Estimate node processing time based on historical data and node characteristics
37.             Apply load balancing algorithm to assign nodes to workers considering:
38.                 Current worker utilization and capacity
39.                 Estimated processing time for candidate nodes
40.                 Worker-specific performance history
41.             Apply load control algorithm to adjust assignments:
42.                 Monitor worker completion times from previous iterations
43.                 Redistribute nodes from overloaded to underutilized workers
44.                 Balance workload to minimize synchronization wait times
45.         Master and workers perform parallel node selection:
46.             Select nodes from assigned private queues
47.             Evaluate nodes and handle pruning with local domain propagation
48.             Perform separation **if** node remains active
49.             Store basis and prepare for next dive iteration
50.         Collect unselected nodes from private queues back to global queue
51.         Apply serial fallback selection for workers without assigned nodes

| | |
|---|---|
| 52. | Update active worker list based on successful node assignments |
| 53. | Record timing statistics and load balancing effectiveness metrics |
| 54. | Terminate worker threads and clean up parallel resources |
| 55. | Return optimal solution with performance statistics |